\documentclass[fleqn,10pt]{wlscirep}
\usepackage[utf8]{inputenc}
\usepackage[T1]{fontenc}
\usepackage{multirow}
\usepackage{multicol}
\usepackage{makecell}
\usepackage{rotating}

\newcommand{\etal}{\textit{et al}.}
\newcommand{\ie}{\textit{i}.\textit{e}.}

\title{Deep segmentation networks predict survival of non-small cell lung cancer}

\author[1,2,3,+]{Stephen Baek}
\author[1,+]{Yusen He}
\author[2]{Bryan G. Allen}
\author[2]{John M. Buatti}
\author[5]{Brian J. Smith}
\author[4]{Ling Tong}
\author[1]{Zhiyu Sun}

\author[6]{Jia Wu}
\author[6]{Maximilian Diehn}
\author[6]{Billy W. Loo}

\author[2]{Kristin A. Plichta}
\author[2]{Steven N. Seyedin}
\author[2]{Maggie Gannon}
\author[2]{Katherine R. Cabel}
\author[2,*]{Yusung Kim}
\author[2,3,*]{Xiaodong Wu}
\affil[1]{University of Iowa, Department of Industrial and Systems Engineering, Iowa City, IA 52242, United States}
\affil[2]{University of Iowa, Department of Radiation Oncology, Iowa City, IA 52242, United States}
\affil[3]{University of Iowa, Department of Electrical and Computer Engineering, Iowa City, IA 52242, United States}
\affil[4]{University of Iowa, Department of Business Analytics, Iowa City, IA 52242, United States}
\affil[5]{University of Iowa, Department of Biostatistics, Iowa City, IA 52242, United States}
\affil[6]{Stanford University, Stanford Cancer Institute, Palo Alto, CA 94304, United States}
\affil[*]{\{yusung-kim, xiaodong-wu\}@uiowa.edu}

\affil[+]{these authors contributed equally to this work}

\begin{abstract}
Non-small-cell lung cancer (NSCLC) represents approximately 80-85\% of lung cancer diagnoses and is the leading cause of cancer-related death worldwide. Recent studies indicate that image-based radiomics features from positron emission tomography/computed tomography (PET/CT) images have predictive power for NSCLC outcomes. To this end, easily calculated functional features such as the maximum and the mean of standard uptake value (SUV) and total lesion glycolysis (TLG) are most commonly used for NSCLC prognostication, but their prognostic value remains controversial. Meanwhile, convolutional neural networks (CNN) are rapidly emerging as a new method for cancer image analysis, with significantly enhanced predictive power compared to hand-crafted radiomics features. Here we show that CNNs trained to perform the tumor segmentation task, with no other information than physician contours, identify a rich set of survival-related image features with remarkable prognostic value. In a retrospective study on pre-treatment PET-CT images of 96 NSCLC patients before stereotactic-body radiotherapy (SBRT), we found that the CNN segmentation algorithm (U-Net) trained for tumor segmentation in PET and CT images, contained features having strong correlation with 2- and 5-year overall and disease-specific survivals. The U-Net algorithm has not seen any other clinical information (e.g. survival, age, smoking history, etc.) than the images and the corresponding tumor contours provided by physicians. In addition, we observed the same trend by validating the U-Net features against an extramural data set provided by Stanford Cancer Institute. Furthermore, through visualization of the U-Net, we also found convincing evidence that the regions of metastasis and recurrence appear to match with the regions where the U-Net features identified patterns that predicted higher likelihoods of death. We anticipate our findings will be a starting point for more sophisticated non-intrusive patient specific cancer prognosis determination. For example, the deep learned PET/CT features can not only predict survival but also visualize high-risk regions within or adjacent to the primary tumor and hence potentially impact therapeutic outcomes by optimal selection of therapeutic strategy or first-line therapy adjustment.
\end{abstract}
\begin{document}

\flushbottom
\maketitle
 \thispagestyle{empty}

% \noindent Key words: {PET/CT}; Cancer survival; Radiomics; Prognosis; Deep learning.

\section*{Introduction}

According to the World Health Organization (WHO), lung cancer remains the leading cause of cancer-related deaths worldwide, with 2.1 million new cases diagnosed and 1.8 million deaths in 2018\cite{WHO2018}. NSCLC accounts for 80-85\% of lung cancer diagnoses\cite{ACSNSCLC} and the five-year survival rate of NSCLC remains low (23\%) compared to other leading cancer sites such as colorectal (64.5\%), breast (89.6\%), and prostate (98.2\%)\cite{cancerstat}. Historically, the tumor, nodes, and metastases (TNM) staging system has served as the major prognostic factor in predicting therapeutic outcomes, but it does not differentiate responders and non-responders within the same stage\cite{Woodard2016}. The maximum and the mean of standard uptake values ($\text{SUV}_\text{MAX}$ and $\text{SUV}_\text{MEAN}$) have been reported for their correlation with survival\cite{BERGHMANS20086,PAESMANS2010612,BOLLINENI2012e551}, but are of limited clinical value due to their unsatisfactory predictive power and lack of robustness\cite{BURDICK20101033,Agarwal2010}. Other prognostic markers have also been studied, including TLG, which incorporates metabolic tumor volume (MTV) and metabolic activity (TLG = MTV $\times$ $\text{SUV}_\text{MEAN}$). Reports\cite{chen2012prognostic,ZAIZEN20124179,MEHTA2014268} suggest that TLG may have better prognostic power than $\text{SUV}_\text{MAX}$ or $\text{SUV}_\text{MEAN}$. These metrics, however, are not optimal and do not provide a comprehensive image-based analysis of tumors\cite{Chicklore2013}. More recently, radiomics approaches, which employ semi-automated analysis based on a few hand-crafted imaging features describing intratumoral heterogeneity, demonstrated higher prognostic power\cite{LEE2017297,10.1371/journal.pone.0192859}. However, these features still have limited predictive power ranging between 0.5 and 0.79 in terms of the area under the ROC (AUC)\cite{10.1371/journal.pone.0192859,Fried2016,zhang2017}. Recent literature on CNNs demonstrates their strong potential for cancer prognostication\cite{Paul2016,diamant2019}, however the clinical implications of deep learning remain questioned due to the limited interpretability of CNNs.

\begin{figure}[t]
\centering
\includegraphics[width=\linewidth]{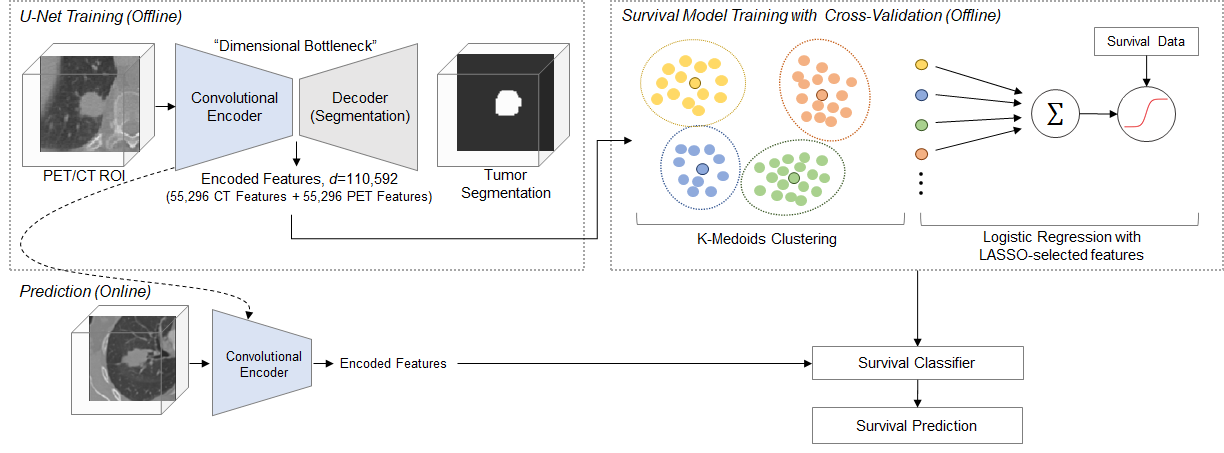}
\caption{ \textbf{Schematic diagram of the survival prediction framework.} The proposed framework consists of two major components: the U-Net segmentation network and the survival prediction model. The U-Net is trained with PET/CT images and corresponding physician contours but without survival-related information. The “dimensional bottleneck” at the middle of the U-Net produces latent variables summarizing image features (55,296 features from CT + 55,296 features from PET), which we hypothesize to be potentially relevant to cancer survival. These features are then clustered by $k$-medoids clustering approach in an unsupervised manner. Next, the LASSO method is used to select medoid features from the clusters based on their associations with survival. Last, a logistic regression model is trained for survival prediction, and survival prediction is performed when a new patient arrives with features extracted from the same U-Net.}
\label{fig:schematic_diagram}
\end{figure}

Here, we propose an interpretable and highly accurate framework to solve this problem by capitalizing on the unprecedented success of deep convolutional neural networks (CNN). More specifically, we investigate U-Net\cite{ronneberger2015u}, a convolutional encoder-decoder network that has demonstrated exceptional performance in tumor detection and segmentation tasks. Illustrated in Fig,~\ref{fig:schematic_diagram}a, U-Net takes a three-dimensional (3D) volume image as an input, processes it through a ``bottleneck layer'' where the image features are compressed, and reconstructs the image into a binary segmentation map indicating a pixel-wise tumor classification result. Here, we focused on the information encoded at the bottleneck layer, which contains rich visual characteristics of the tumor and hypothesized that the encoded information at this layer might be relevant to the tumor malignancy, and thus cancer survival, which is the central hypothesis of this paper.

\section*{Results}
In prior studies\cite{wu2018multi,Zhong2019}, we analyzed PET/CT images of 96 NSCLC patients that were obtained within 3 months prior to SBRT, whose summary statistics are illustrated in Extended Data Fig.~\ref{fig:Summary_stat} and Extended Data Table~\ref{tbl:summary_stat}. For each volume image, the region of interest (ROI) with a dimension of 96 mm $\times$ 96 mm $\times$ 48 mm was set around each SBRT treated tumor location and the image was cropped to the ROI volume. Two separate U-Net models were trained to perform tumor segmentation in PET and CT images, respectively. Each of the models was trained with 38 ROI images and the corresponding physician contours, but no other information such as survival time was provided. Segmentation performance was tested on 22 independent ROI images that were not included in training, and the average S{\o}rensen-Dice similarity coefficients (DSC) were $0.861 \pm 0.037$ (mean $\pm$ std) and $0.828 \pm 0.087$ for CT and PET, respectively. After training, each U-Net model learned to encode 55,296 features at the bottleneck layer for each patient, resulting in a total of 110,592 features per patient.

% \begin{figure}[ht]
% \centering
% \includegraphics[width=0.7\linewidth]{img/summary_stat.png}
% \caption{Summary statistics of the data set used in this study.}
% \label{fig:summary_stat}
% \end{figure}

These features are an intermediate throughput of U-Net, and are then decoded to generate an automated segmentation in the network. It is likely that these features summarize some rich structural and functional geometry of the intratumoral and peritumoral area, some of which might be relevant to cancer survival.  We test this proposition by conducting validation studies and examining the statistical prognostic power of those features. One challenge here is that the number of features ($d=110,592$) is substantially larger than the number of cases, making a statistical analysis prone to overfitting. To this end, we first reduce the number of features via an unsupervised feature selection process in which the survival information is hidden. The $k$-medoids clustering method \cite{park2009simple} is employed to serve this purpose, as the method is known to be able to select representative features from a large pool of inter-correlated features in similar literature \cite{uthoff2019machine}. For the feature similarity measure in $k$-medoids clustering, we use the Pearson correlation distance defined as $1-R$, where $R$ is the Pearson correlation coefficient. The optimal number of clusters is determined by the Silhouette method \cite{rousseeuw1987silhouettes}. Finally, given the optimal $k$ clusters, the medoids of the clusters are selected to be the representative features of the clusters and form an unsupervised feature pool. We choose the final set of features via the least absolute shrinkage and selection operator (LASSO)\cite{tibshirani1996regression} on the preliminary feature pool obtained from the $k$-medoids method. LASSO attempts to select a few features among the cluster centers that have a strong relationship with the survival outcomes using the fixed regularization parameter. The selected features then serve as regressors in the logistic regression model to predict the cancer survival outcome. Since the performance of $k$-medoids clustering is also dependent on a random initialization of the cluster centers, we tested 10 times for each $k$ and computed the mean of the summation of the inner cluster distances to ensure consistency of the clustering outcome. We then computed the curve of the Silhouette value with respect to the number of clusters in total to determine the optimal number of clusters. The medoids of the feature clusters are selected to form a reduced feature set.

% \begin{figure}[ht]
% \centering
% \includegraphics[width=0.8\linewidth]{img/pvalue_histogram.png}
% \caption{\textbf{Feature selection.} (\textbf{a}) Log-scaled histogram of $p$-values for each survival category. Roughly 2,000 features in each category have $p$-value less than 0.05. (\textbf{b}) Intersection between different survival categories. `NS' indicates not significant features. Black dots are features that were significant ($p<0.05$) in all four categories. There were total 139 and 292 such features in CT and PET, respectively, but only 10 of them with the smallest $p$-values in all four categories were plotted. (\textbf{c}) Distribution of $p$-values among LASSO-selected features and across the others. Dashed line is the cutoff value ($p=0.05$).}
% \label{fig:pvalue_histogram}
% \end{figure}

We then examined the selected features and their prognostic power via cross-validation on the NSCLC data set collected at the University of Iowa Hospitals and Clinics (UIHC). As summarized in Extended Data Fig~\ref{fig:Summary_stat} and Extended Data Table~\ref{tbl:summary_stat}, the UIHC data set is comprised of primary and follow-up PET/CT images of total 96 NSCLC cases with their survival status and other clinical meta data. Information on the cause of death is also available so that the deceased cases can be further broken down to overall and disease-specific deaths. On this data set, we aim to predict four survival categories, namely, 2-year overall survival (2OS), 5-year overall survival (5OS), 2-year disease-specific survival (2DS), and 5-year disease-specific survival (5DS). Total $N=96$ cases qualify for 2OS category, 74 for 5OS, 92 for 2DS, and 45 for 5DS, depending on the survival status and the cause of death. Other clinical meta data, such as sex, age, smoking history, and tumor, lymph nodes, and metastasis (TNM) staging, exist in the data set and may provide an improved prognostic power when added as regressors. However, we exclude all other parameters but U-Net encoded image features, in order to focus the analysis on the image features only. Only primary images are encoded via U-Net and used for prediction. Follow-up images are reserved for comparison and discussion later in this paper. Each cross validation experiment is comprised of unsupervised feature selection using $k$-medoids clustering, LASSO-based feature selection, and training of the logistic regression model. These tasks are performed independently from the other cross-validation experiments. In each experiment, we measure the accuracy, sensitivity, specificity, and the area under the receiver operating characteristic curve (AUC) per each survival category. We compute the average and the standard deviation of these performance metrics across the cross-validation experiments to derive the estimated performance metrics and their 95\% confidence intervals (95\% CI).

\begin{figure}[t]
\centering
\includegraphics[width=\linewidth]{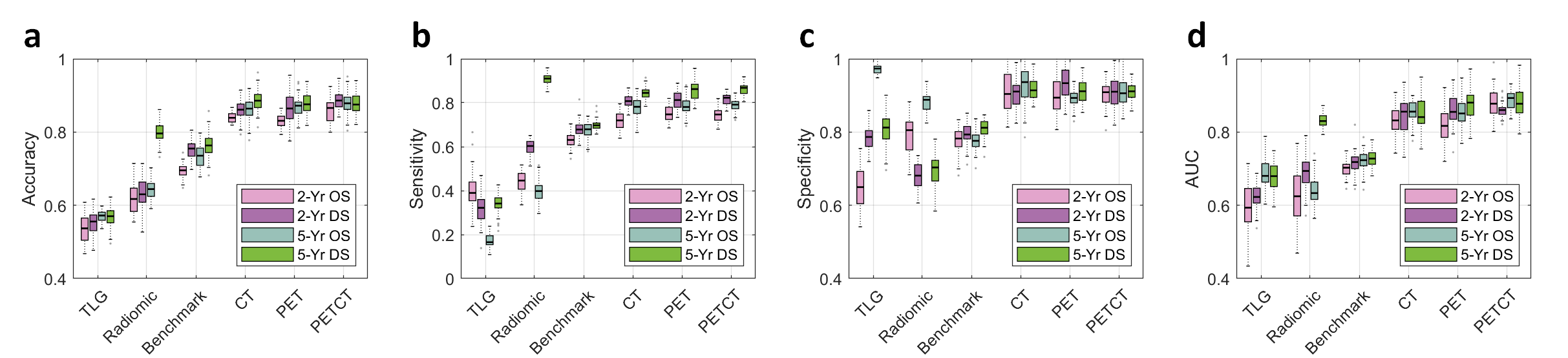}
\caption{\textbf{Prognostic performance of the U-Net features.} There are four survival categories being tested: 2-year overall survival (2-yr. OS), 5-year overall survival (5-yr. OS), 2-year disease-specific survival (2-yr. DS), and 5-year disease-specific survival (5-yr. DS). The U-Net features are compared against the conventional TLG metric, the 17 radiomics features defined in Oikonomou \etal\cite{oikonomou2018} and the benchmark CNN prediction model in Hosny \etal\cite{hosny2018deep}. The box plots represent the average performance scores as indicted by the central mark and 25th and 75th percentiles across 6-fold cross validation experiments. (\textbf{a}) Overall prediction accuracy (proportion of the correct prediction over the entire data set). (\textbf{b}) Sensitivity (correct prediction of death over all death cases). (\textbf{c}) Specificity (correct prediction of survival across all survival cases). (\textbf{d}) AUC of the receiver operating characteristics (ROC) curve.}
\label{fig:accuracy}
\end{figure}

In such cross-validation experiments, estimated AUC of the proposed prediction model is 0.88 (95\% CI: 0.80-0.96) for the prediction of 2OS. For other survival criteria, the estimated AUCs are similar, namely 0.89 (95\% CI: 0.85-0.93) for 5OS, 0.86 (95\% CI: 0.81-0.91) for 2DS, and 0.88 (95\% CI: 0.81-0.95) for 5DS. Note that the estimated AUC values of the conventional TLG and other radiomics markers \cite{oikonomou2018} range between 0.60 and 0.83 on the same data set. A more recent deep learning method reported in Hosny \etal\cite{hosny2018deep} produces AUCs between 0.70$\sim$0.73 on the same data set. A graphical illustration of the result as well as the full set of performance metrics are available in Fig.~\ref{fig:accuracy} and Extended Data Table~\ref{tbl:performance}, respectively.

\begin{figure}
\centering
\includegraphics[width=\linewidth]{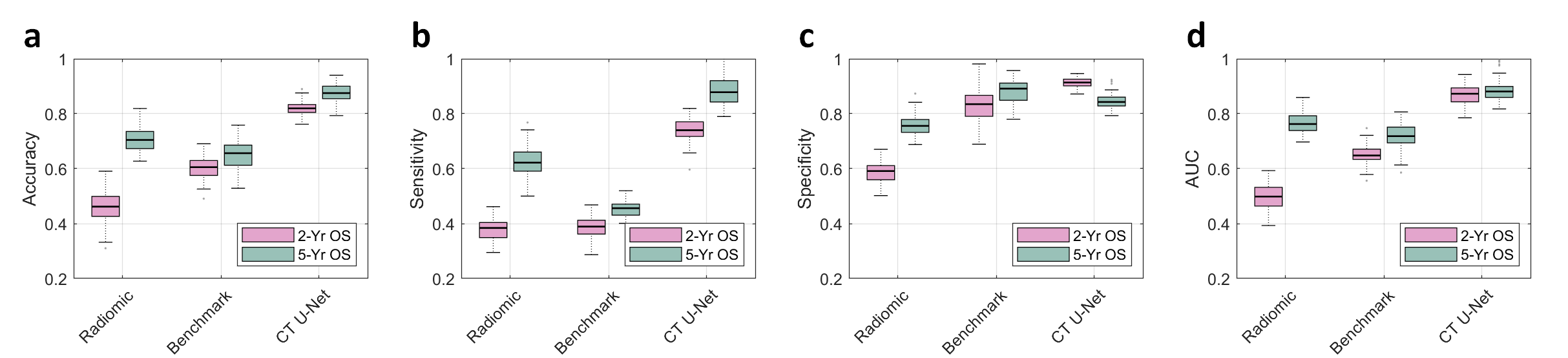}
\caption {\textbf{Prognostic performance on an extramural data set.}  The extramural data set provided by Stanford only includes CT images and not PET images. Additionally, the disease-specific survival information is not provided. Therefore, in this experiment, two survival categories are being tested: 2-year overall survival (2-yr. OS) and 5-year overall survival (5-yr. OS). The U-Net features are compared against the 6 CT-based radiomics features (Radiomic) defined in Oikonomou \etal\cite{oikonomou2018} and the benchmark CNN prediction model (Benchmark) in Hosny \etal \cite{hosny2018deep}. The box plots represent the average performance scores as indicted by the central mark and 25th and 75th percentiles across experiments tested on extramural Stanford data set. (\textbf{a}) Overall prediction accuracy (proportion of the correct prediction over the entire data set). (\textbf{b}) Sensitivity (correct prediction of death over all death cases). (\textbf{c}) Specificity (correct prediction of survival over all survival cases). (\textbf{d}) AUC of the receiver operating characteristics (ROC) curve.}
\label{fig:stanford}
\end{figure}

Moreover, we further validated the result on an extramural dataset provided by the Stanford Cancer Institute. The Stanford data set is comprised of primary CT images of 26 NSCLC cases which received SBRT, of which 18 survived and 8 died according to 2 year OS and 1 survived and 25 died according to 5 year OS. Neither PET images nor disease-specific survival information were available in the Stanford data set. Training of the U-Net, feature selection ($k$-medoids and LASSO), and construction of the survival model are performed only on the UIHC data set, and the Stanford data set is used for validation only. In this setting, the estimated AUC is 0.87 (95\% CI: 0.80-0.94) for 2 year OS, and 0.90 (95\% CI: 0.82-0.98) for 5 year OS. More detailed results are illustrated in Fig.~\ref{fig:stanford} and Extended Data Table~\ref{tbl:stanford_stat}.

\begin{figure}[t]
\centering
\includegraphics[width=0.5\linewidth]{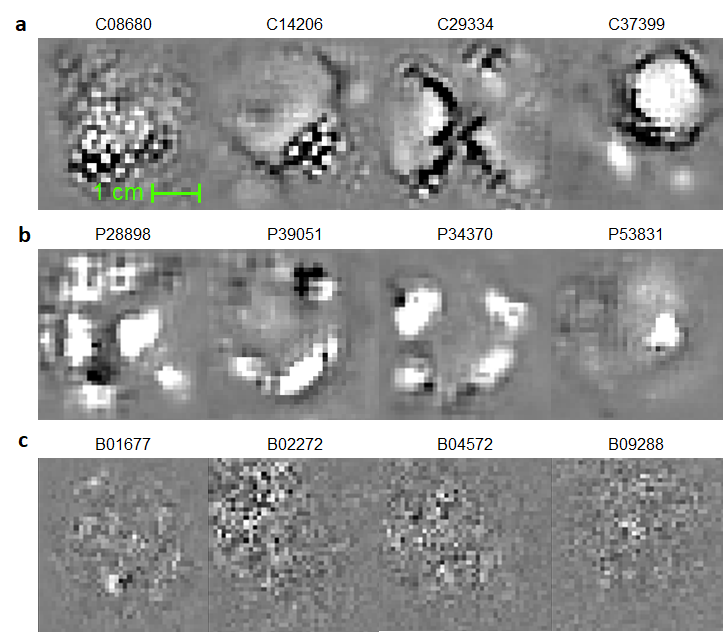}
\caption{\textbf{Survival-related features captured by the U-Nets.}
During training, CNNs essentially learn ``templates/patterns'' from training images and apply these templates to analyze and understand images. (\textbf{a}) Image templates that the U-Nets have captured for the segmentation task, in CT and (\textbf{b}) in PET. Note these templates are learned in an unsupervised manner, without any survival-related information provided, despite which these were later discovered to be survival-related. Note that the templates captured by U-Net are characterized by their sensical and interpretable geometric structures . For example, C37399 appears to be a template looking for a tumor-like shape at the top-right corner and a tube-like structure at the bottom-left. In addition, C08680 appears to look for a textural feature of the tumor. (\textbf{c}) In contrast, image templates learned by direct fitting of a CNN model to the survival data\cite{hosny2018deep}. Note the features in (\textbf{c}) are less interpretable compared to the U-Net features.}
\label{fig:activation_maximization}
\end{figure}

Meanwhile, we visualized the features learned by U-Net to develop an intuitive understanding of what those prognostic markers represent. The image features learned by U-Net are essentially artificial neurons in deep neural networks. In principle, we can visualize a neuron by showing multitudes of image patterns and observing which image pattern activates the neuron the most. Practically, we employ an optimization-based approach\cite{yosinski2015} where the objective is to maximize an individual neuron's activation value by manipulating the input image pattern:
\begin{equation}
    \mathbf{X}^*= \arg\max_{\mathbf{X}} q(\mathbf{X} | \mathbf{W}, \mathbf{b}),
    \label{eqn:activation}
\end{equation}
where $q( \cdot | \mathbf{W}, \mathbf{b})$ is the U-Net encoder with the trained model parameters $\mathbf{W}$ and $\mathbf{b}$, and $\mathbf{X}$ is the input image pattern. Displayed in Fig.~\ref{fig:activation_maximization} and Extended Data Fig.~\ref{fig:activation_maximization_full} are visualizations of features captured by artificial neurons.

\begin{figure}[!]
\centering
\includegraphics[width=0.6\linewidth]{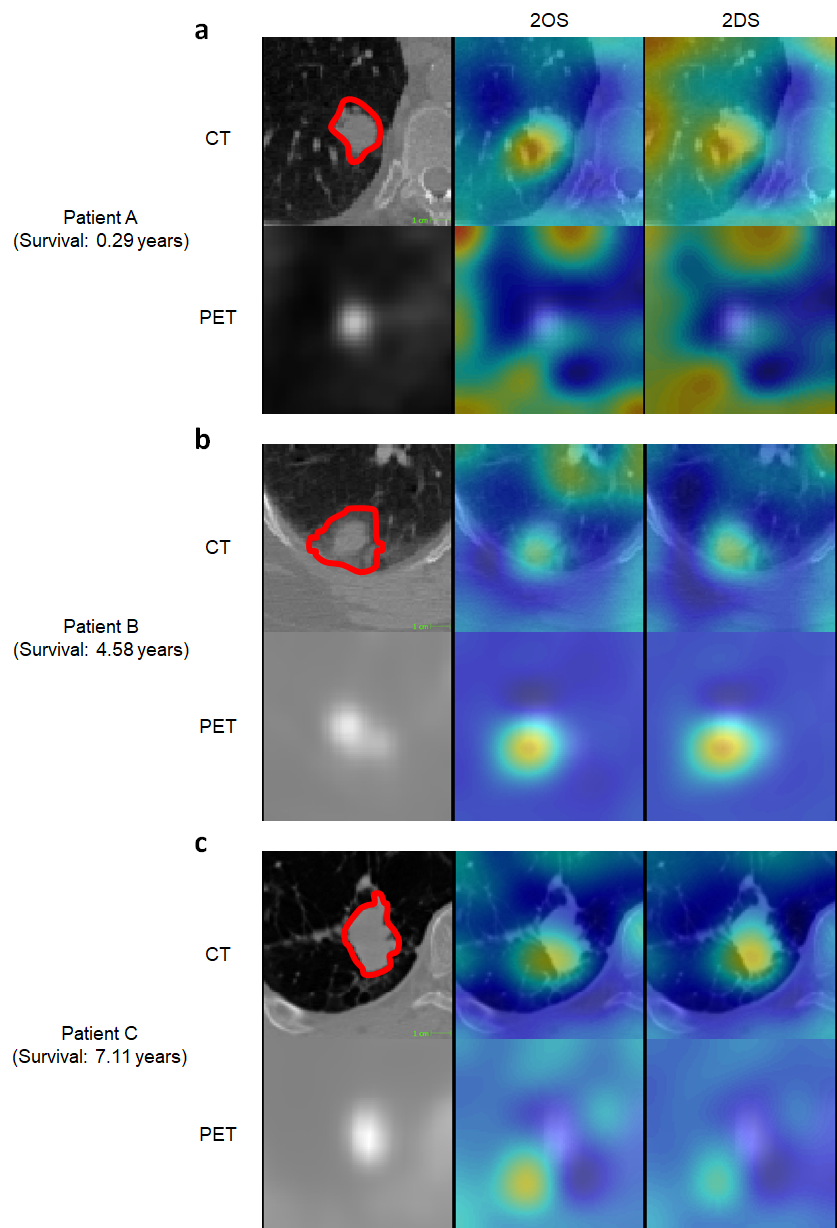}
\caption{\textbf{Visualization of the U-Net features.} Regions that predicted death of the patients obtained via a guided backpropagation method\cite{selvaraju2017}. Trivially, tumoral regions are highlighted in red in the heatmap. However, some of the heated regions outside of the tumoral volume matched with the actual locations of recurrences and metastases when they were compared with the post-therapeutic images and clinical records, rendering a great potential as a practical, clinical tool for patient-tailored treatment planning in the future. (\textbf{a}) Patient deceased in 0.29 years after the acquisition of the images. (\textbf{b}) Deceased after 4.58 years. (\textbf{c}) Deceased after 7.11 years.}
\label{fig:heatmap}
\end{figure}

We also visualized which regions in the patient images predicted low survival probability. We employed a guided gradient backpropagation approach\cite{selvaraju2017}. The main idea of the guided backpropagation algorithm is to compute $\frac{\partial P}{\partial x_{i,j,k}}$ where $P$ is the probability of death and $x_{i,j,k}$ is a voxel value at the position $(i,j,k)$ in the patient image. The gradient $\frac{\partial P}{\partial x_{i,j,k}}$ can be interpreted as the change of the death probability when the voxel $x_{i,j,k}$ changes to a different value. If the voxel was not so significant in predicting death, the gradient value would be small, where as if the voxel played an important role in predicting high probability of death, the gradient value would be greater. Displayed in Fig.~\ref{fig:heatmap} and Extended Data Figs.~\ref{fig:riskmap_under2}-\ref{fig:riskmap_over5} are heat maps representing the gradient. Heated regions (red) are the areas that lowered the probability of survival whereas the other areas (blue) are the ones that had negligible effect on the survival.

% \begin{figure}[!]
% \centering
% \includegraphics[width=0.65\linewidth]{img/visualization.png}
% \caption{\textbf{Visualization of the U-Net features.} (\textbf{a}) Image patterns that are captured by each survival-related U-Net feature. During training, CNNs essentially learn ``templates'' from training images and apply these templates to analyze and understand images. Displayed here are some of these templates that the U-Nets have captured for the segmentation task originally, but discovered in our study to be relevant to cancer survival. For example, C37399 appears to be a template looking for a tumor-like shape at the top-right corner and a tube-like structure at the bottom-left. In addition, C08680 appears to look for a textural feature of the tumor. (\textbf{b}) Regions that predicted death of the patients obtained via a guided backpropagation method\cite{selvaraju2017}. Trivially, tumoral regions are highlighted in red in the heatmap. However, some of the heated regions outside of the tumoral volume matched with the actual locations of recurrences and metastases when they were compared with the post-therapeutic images and clinical records, rendering a great potential as a practical, clinical tool for patient-tailored treatment planning in the future.}
% \label{fig:visualization}
% \end{figure}

\section*{Discussion}

As illustrated Fig.~\ref{fig:accuracy} and Extended Data Table~\ref{tbl:performance}, the contrast in the predictive performance between the U-Net features and the conventional imaging features (SUV$_{MEAN}$, SUV$_{MAX}$ and TLG) was evident, quantitatively proving the strong prognostic power of the U-Net features. There also was a noticeable enhancement of prognostic performance when the U-Net features were compared with a recent deep learning approach as in Hosny \etal\cite{hosny2018deep} The same trend could be observed in Fig.~\ref{fig:stanford} and Extended Data Table~\ref{tbl:stanford_stat} where we validated the U-Net features against an extramural data set, confirming the enhanced prognostic performance of the U-Net features. Here, it is worth reemphasizing that the U-Net was trained without any survival-related information, and, thus it is highly unlikely that the U-Net-learned features were overfitted to the survival data or biased towards them. Nonetheless, while these U-Net features were identified independently from the survival data, the U-Net features demonstrated strong evidence of correlation and thus prognostic power for NSCLC survival as discussed above.

Here, a justification should be necessary on the rationale behind taking a ``detour'' by training a segmentation network first, then extracting prognostic image features, and training a survival model, as opposed to directly training a CNN model to the survival data as in other literature.\cite{hosny2018deep, paul2016deep, yao2016imaging} To this end, we make the following arguments. First, a CNN model trained on a segmentation task is more robust to overfitting and is more generalizable, as the image features themselves are developed from an unsupervised training. A direct prediction of survival from an image tends to be less robust and less generalizable, as the millions of parameters in a CNN can easily misconstrue the trend, and this process is difficult to control. Secondly, segmentation-trained features contain more structural, geometric features that are human-interpretable, whereas direct-trained features tend to be biased towards less-intuitive and fuzzy texture patterns. It is, in fact, well-understood now that CNNs tend to be biased towards texture rather than shape, while the human cognitive process works in the opposite.\cite{kubilius2016deep} For example, Geirhos et al.\cite{geirhos2018imagenet} conducted an experiment where the texture of a cat in an image was replaced by the texture of an elephant---CNNs started to discern the image as an elephant while humans still categorized the image as a cat. From this, we can infer that the direct training of a CNN for survival prediction would result in a development of less sensical features lacking geometric information. In contrast, segmentation-trained features are incentivized to focus on geometric structures and shapes rather than textures, and thus are more sensical.

The above arguments are indeed supported by the visualization results in Fig.~\ref{fig:activation_maximization} and Extended Data Fig.~\ref{fig:activation_maximization_full}, where the benchmark CNN method appears to be capturing some noisy texture patterns. In contrast, many of the U-Net features appear to be capturing more sensical and interpretable geometric shapes from the image, such as tumor-like blobs (e.g. C00048, C25988, P39051, P47258) or heterogeneity of the tumor (e.g. C08680). Interestingly, some of the U-Net features, for example C01777 and C37399, were looking for tube-like structures nearby the tumor-like blobs, which might be capturing blood vessels and lymphatics in the peritumoral area. This is consistent with the widely accepted clinical knowledge that tumors can show enhanced growth in the presence of nearby vessels and lymphatics as they carry nutrition to supply the tumoral growth.

\begin{figure}[!]
\centering
\includegraphics[width=\linewidth]{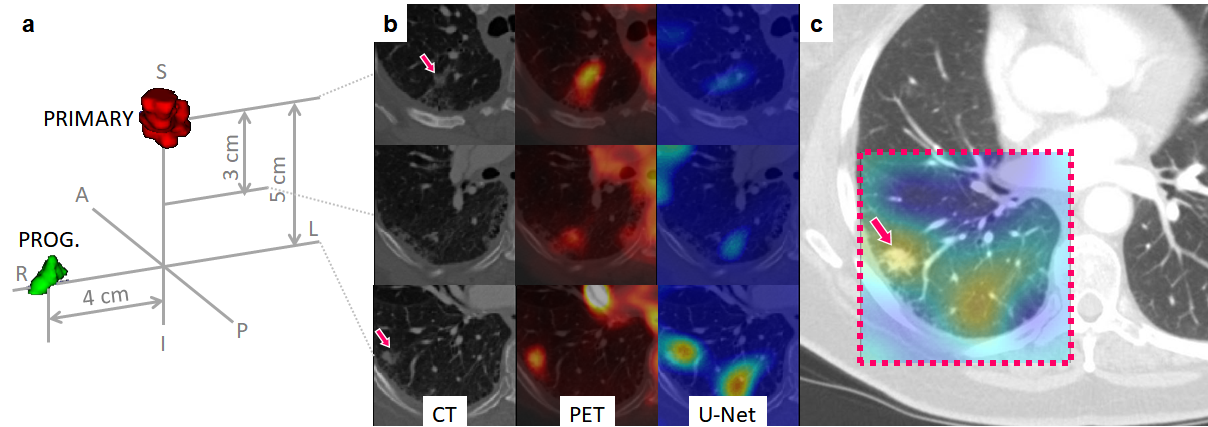}
\caption{\textbf{Correlation between U-Net visualization and cancer progression.} Post-SBRT CT images were compared with the U-Net visualization results. We observed an agreement of the heated regions with the actual location of recurrence as in this example. (\textbf{a}) A 3D rendering showing the location of the primary tumor volume (red) and the progression region (green) of the case IA001765; (\textbf{b}) Pre-SBRT transversal slices at the primary tumor location (top), 3 centimeters below (middle row), and 5 centimeters below (inferior row); (\textbf{c}) A follow up (post-SBRT) image of the same patient. The dashed box indicates the estimated corresponding ROI to the primary (pre-SBRT) CT slices. The heat map generated based on pre-SBRT, \ie, the same heat map as in the bottom row of (\textbf{b}), is superimposed on top of the ROI on the post-SBRT image. Notice that the recurrence location coincides with the heated area.}
%Axial slice from a primary (pre-SBRT) CT image of a patient case IA001765. The slice is inferior to the primary target volume of SBRT so the primary tumor is not visible. (\textbf{b}) Corresponding U-Net visualization overlaid on pre-SBRT CT images. There are two highlighted regions in the heat map: One is a small nodule that was not treated SBRT and another is the primary tumor (\textbf{c}) Post-SBRT CT image of the same patient. Dashed box indicates the estimated corresponding ROI to the primary CT slices.{\color{red}(\textbf{d-e}) are risk map visualization of adjacent slices (0.5 and 1 cm) inferior to the tumor recurrence}. (\textbf{f-g}) are risk map visualization of adjacent slices (0.5 and 1 cm) superior to the tumor recurrence}. The heat map in (\textbf{b}) is superimposed to the ROI. Notice that the recurrence location coincides with the heated area.}
\label{fig:progression}
\end{figure}

Meanwhile, we further visualized the proposed prediction model by creating a heat map highlighting regions that predicted low survival probability. The map generated by the guided back-propagation algorithm confirmed that the prediction model was looking at clinically sound regions for making predictions by responding more to tumor and surrounding tumor regions than other regions. In addition to the results presented above, this is yet more evident that the prediction model has produced generalizable features and rules for making prognostic prediction, and is thus not overfitted. Moreover, through comparison with post-SBRT CT images and clinical records of the patients, we observe that the heat map has the potential to identify regions of progressions or recurrence. For the case illustrated in Fig.~\ref{fig:progression} as an example, a small nodule had been found initially at the progression region, but was not treated by SBRT as the region was not originally identified as tumor by PET. However, this region was identified as a high risk region on heat map by the U-Net as well as the primary tumor region of SBRT. Tumor recurrence was found on post-SBRT CT images and matched the highlighted risk region on pre-SBRT U-Net generated heat map. From such observations, the heat map visualization has the potential to identify regions at high risk for tumor progression or recurrence that could be utilized for the purpose of assisting patient-tailored treatment planning in the future. For this reason, we believe this indicates more rigorous risk map developments and requires quantitative follow-up validations, which will be our subsequent project.

In summary, we discovered that the U-Net segmentation algorithm trained for automated tumor segmentation on PET/CT, codifies rich structural and functional geometry at the bottleneck layer. These codified features, in turn, could be used for survival prediction in cancer patients even though the U-Net was trained without any survival-related information. The survival model based on such U-Net features demonstrated significantly higher predictive power compared to conventional PET-based, metabolic burden metrics such as TLG or relatively recent hand-crafted radiomics approaches. The validity of this discovery was further confirmed by the validation on an extramural data set provided by the Stanford Cancer Institute. Furthermore, we visualized the survival-related U-Net features and observed that they were indeed depicting intratumoral and/or peritumoral structures that had been previously acknowledged as potentially relevant to cancer survival. Our approach awaits a further validation against a larger number of observations and in a larger variety of cancer types. However, there was not enough clinical evidence to conclude that the visualization of the U-Net features may identify potential regions of recurrence and metastasis and, thus, a follow-up study is suggested. Our findings may suggest a new starting point for quantitative image-based cancer prognosis with a great deal of important new knowledge to be discovered.

\section*{Methods}

\textbf{Subjects}~~~~~PET-CT images of NSCLC patients who received SBRT at the University of Iowa Hospitals and Clinics were investigated in this study. The images were obtained using a dual PET-CT scanner (Biograph 40 PET/CT, Siemens Medical Solutions USA, Inc., Hoffman Estates, IL). CT image used for SBRT planning were co-registered with CT images of PET-CT datasets using a deformable image registration (DIR). Using the deformation map (vectors), PET images were resampled with the primary CT images of SBRT plan. Post-SBRT follow-up images were used only for qualitative validation of the visualization results. The gross tumor volume (GTV) for each of the images was delineated by radiation oncologists on both CT and PET images with the guidance of the corresponding images in the other modality. All tumor contouring was completed using VelocityAI (Varian Medical System, Inc., Palo Alto, CA).

A total of 96 cases (Male=44; Feamle=52) were investigated in this study. Patients' group stages vary in sub-categories including 41 in stage I, 10 in stage II, 16 in stage III, 29 in stage IV. Meanwhile, the histology was confirmed by a thoracic pathologist based off visual interrogation of the biopsied or surgically resected specimen. Histologies among the 96 patients include 48 adenocarcinoma, 41 squamous cell, 1 adenosquamous, 3 metastases from previous NSCLC, and 1 without biopsy. For overall survival from SBRT, the 2-year survival rate is 54\% and the 5-year survival rate is 6\%. For overall survival from diagnosis, the 2-year survival rate is 66\% and the 5-year survival rate is 28\%. For disease-specific survival from SBRT, the 2-year survival rate is 61\% and the 5-year survival rate is 23\%. For disease-specific survival from diagnosis, the 2-year disease specific survival rate is 73\% and the 5-year disease specific survival rate is 51\%. Among a total 96 patients, the qualified portion utilized in this research in each survival sub-category has been visualized and discussed in Extended Data Fig.~\ref{fig:Summary_stat} respectively.

In this study, the total of 96 utilized NSCLC patients were retrospectively analyzed after approval from the University of Iowa Institutional Review Board (IRB: 200503706; Name: The utility of imaging in cancer: staging, assessment of treatment response, and surveillance). All data collection, experimental procedures, and methods applied are in accordance with the relevant guidelines and regulations. All patients consented for the use of their clinical information and medical images. All participants enrolled in this study signed an informed consent developed and approved by the University of Iowa Institutional Review Board. All scans were in digital imaging and communications in medicine (DICOM) format and de-codified.

\noindent\textbf{Data Processing}~~~~~Each pair of the co-registered PET-CT images were resampled with an isotropic spacing in all three dimensions. Each image was then cropped into a fixed size of $96 \times 96 \times 48$ voxels where each voxel represents 1 cubic millimeter (mm$^3$) after resampling. Within the resampled voxels, intensity values were clipped to the range of [-500, 200] for all CTs and [0.01, 20] for all PETs, to remove outliers. More details are described in our previous work\cite{wu2018multi,Zhong2019}. 

\noindent\textbf{U-Net Features}~~~~~In our previous work\cite{Zhong2019}, two independent 3D U-Net networks were constructed and trained for automated tumor segmentation in PET and CT images, respectively. The U-Net networks comprise two major components: the encoder network and the decoder network (Fig.~\ref{fig:schematic_diagram}). The encoder network takes a $96 \times 96 \times 48$ volume image as an input. The first convolution layer produces 32 features attached to each voxel, representing low-level visual cues such as lines, edges, and blobs. These features are then down-sampled by half in all three dimensions and the down-sampled volume is fed into the second convolution layer, which then produces 64 features per each voxel. This is repeated three more times increasing the number of features to 128, 256, and finally 512, while the volume size is reduced by half in all three dimensions each time. The final $6 \times 6 \times 3 \times 512 (=55,296)$ features that the encoder network produces are an abstract, high-level summary of the input image, which is then decoded by the symmetric decoder to produce the binary segmentation map (1: tumor, 0: none). The convolutional kernels are of a size $3 \times 3 \times 3$ across all layers and the max-pooling layers of a $2 \times 2 \times 2$ window size with a stride of 2 were used for down-sampling. 

 To train the PET-CT segmentation U-Net networks, we utilized co-registered PET-CT scan pairs from 60 patients with primary NSCLC. For each PET-CT image, the slice image size is $512 \times 512$ and the number of slices varies from 112 to 293. The tumor contour on each of the PET and CT scans were labeled by physicians as groundtruth. In data preprocessing, all PET-CT images are resampled with an isotropic spacing of $1 \times 1 \times 1$ in voxels and then cropped at a fixed size of 3D volumes ($96 \times 96 \times 48$) centered on the mass gravity of each tumor. 

All 60 PET-CT scan pairs were split into a training data set with 38 patients and testing data set with 22 patients. Data augmentation was performed using simple translation, rotation and flip operations and the augmented training set has over 3000 3D PET-CT scan pairs respectively.

The 3D-UNet was trained using open source TensorFlow package and ran on NVIDIA GeForce GTX 1080
Ti GPU with 11GB of memory. The Adam optimization method was utilized with a mini-batch size of 1 and for 20 epochs. To prevent overfitting, the weight decay and early-stop techniques were adopted to obtain the best performance on the test set where the DSC value was computed.

After the U-Nets had been trained, PET and CT images of each patient are fed into the U-Nets to produce 55,296 features per imaging modality. These 55,296 features extracted from the U-Net encoder were used for the analyses throughout the paper.  The schematic diagram with respect to the methodologies applied in this research for the feature analysis has been illustrated in Fig.~\ref{fig:schematic_diagram}.

\noindent\textbf{$k$-medoids feature clustering}~~~~~In this research, several $k$-medoids clustering experiments were conducted on the training dataset for the cross-validation experiments. Pearson's correlation distance was employed as the distance metric for clustering the features as expressed in:
\begin{equation}
    \mathcal{D}(\mathbf{X,Y}) =  1- \frac{\text{Cov}(\mathbf{X,Y})}{\sqrt{\text{Var}(\mathbf{X})\text{Var}(\mathbf{Y})}}
\end{equation}
where $X$ and $Y$ denote the two distinct features; $Cov()$ denotes the covariance of the two features and $Var()$ is the variance of the underlying feature. The sum of inner cluster distances were computed by setting various $k$ values, and the optimal number of clusters were determined by the Silhouette method \cite{rousseeuw1987silhouettes}. The medoids of all clusters were selected as candidate features to construct the survival prediction models.

\noindent\textbf{Feature Selection}~~~~~The LASSO regression algorithm was employed to narrow down the scope of analysis to survival-related features from the medoids of clusters obtained from clustering results. The LASSO algorithm used in this study is expressed as:
\begin{equation}
\min_{\beta} \| y - \beta X \|^2_2 + \lambda\|\beta\|_1
\end{equation}
where $y$ denotes the survival outcome (1: alive, 0: dead), $X$ is a vector containing all latent variables extracted from the U-Net network, $\beta$ denotes the coefficient of regression, and $\lambda$ is the penalty coefficient. The L1-norm in the second term penalizes the selection of redundant variables. The parameter $\lambda$ was determined via cross validation on the training dataset. Latent variables that survived the L1-penalty with the best $\lambda$ were selected for the logistic regression model to predict the survival outcome. Using the LASSO-selected variables, we applied logistic regression to estimate the coefficients and predict the survival outcome.

\noindent\textbf{Logistic Regression}~~~~~We formulate survival as a dummy variable. The task of predicting survival outcome can then be formulated as a binary class probability prediction problem and we select the linear logistic regression model as our statistical model:
\begin{equation}
y = \left\{ 1 + e^{-(\beta_0 + \sum_{i=1}^{p}\beta_i x_i)} \right\}^{-1}
\end{equation}
where $y$ denotes the predicted survival probability, $x_i$ is the $i$-th LASSO-selected U-Net feature, and $\beta_i$ is the regression coefficient. The performance of a prediction model was measured via 6-fold cross validation. In experiment, the samples were split into training and test sets. The models were trained with the training set and the test sets were left out for validation. The proportion of survival and death cases were controlled to be equal in the test sets. Reported performance metrics in this paper are based on the statistics of the test set validations across 6-fold cross validation.

\noindent\textbf{Visualization}~~~~~An activation maximization scheme\cite{yosinski2015} was employed to visualize the LASSO-selected U-Net features. For a trained U-Net encoder $q( \cdot | \mathbf{W}, \mathbf{b})$, neurons at the bottleneck layer corresponding to the LASSO-selected features were denoted as $q_i$. Then, Eq.~\ref{eqn:activation} was solved for each individual neuron via gradient ascent:
\begin{equation}
    \mathbf{X}^{(k+1)} = \mathbf{X}^{(k)} + \gamma^{(k)} \nabla q_i (\mathbf{X}^{(k)}),
\end{equation}
where $\mathbf{X}^{(k)}$ is the current solution at $k$-th iteration and $\gamma^{(k)}$ is a step length. We set $\gamma^{(k)}$ as $1/\sigma^{(k)}$ where $\sigma^{(k)}$ denotes the standard deviation of the gradients. The gradient $\nabla q_i$ was computed using the standard backpropagation algorithm. The initial image $\mathbf{X}^{(0)}$ was initialized with random voxel values following the Gaussian distribution {$\mathcal{N}(128,1)$}. Displayed in Fig.~\ref{fig:activation_maximization}a-b are the final solution $\mathbf{X}^*$ after 20 iterations.

Moreover, we also visualized a risk map by evaluating each voxel's contribution to the prediction of survival. We employed a guided backpropagation approach similar to Selvaraju \etal\cite{selvaraju2017}. For each voxel in the input image, with marginal change of the survival probability with respect to the voxel's intensity, defined as $\frac{\partial P}{\partial x_{i,j,k}}$, where $P$ is the probability of death and $x_{i,j,k}$ is a voxel value at the position $(i,j,k)$. In the guided backpropagation process, we rectified the gradient by dropping the negative gradient values to focus on the ``risk''. This was achieved by applying rectified linear unit (ReLU) activation when the values were backpropagated from node to node:
\begin{equation}
    \mathbf{\alpha}^{(m)} =  \max \left( \frac{\partial(P)}{\partial {A}^{(m)}_{i,j,k}},0
    \right)
\end{equation}
where ${A}^{(m)}$ denotes the activation map corresponding to the $m$-th convolutional kernel at the bottleneck encoding. Note that only the LASSO-selected features were involved in the survival model $P$ such that $\frac{\partial(P)}{\partial {A}^{(m)}_{i,j,k}}$ is zero most of the time. Finally, the risk map $\mathcal{R}$ was defined as a linear combination of all activation maps at the bottleneck layer with the coefficients $\mathbf{\alpha}^{(m)}$ obtained from the above:
\begin{equation}
    \mathcal{R}(\mathbf{X}) = \sum_{m}{\alpha}^{(m)}{A}^{(m)}(\mathbf{X}).
\end{equation}

\bibliography{main}

\section*{Acknowledgements}
The authors want to express their gratitude to Dr. Ruijiang Li at Stanford University for his assistance in coordinating the collaboration with Stanford University and for providing the extramural data set. They also thank Dr. Sanjay Aneja at Yale University for providing insight and expertise that greatly assisted the research.
Research reported in this publication was supported by the National Cancer Institute (NCI) of the National Institutes of Health (NIH) under award number 1R21CA209874 and partially by U01CA140206 and P30CA086862.

\section*{Author contributions statement}

Y.K. and X.W. conceived and managed the experiments, S.B., Y.H. and L.T. conducted and analyzed the experiments, Y.H. and Z.S. performed visualization experiments, S.B. and B.J.S. advised the statistical analysis, Y.H., M.G., and K.R.C. processed the raw data, B.G.A., J.M.B., and K.A.P. provided the contours for training of the U-net, B.G.A., J.M.B., and S.N.S. advised clinical discussions. R.L., J.W., M.D., and B.L. provided and processed the extramural data set. All authors reviewed the manuscript.

% \section*{Additional information}

% To include, in this order: \textbf{Accession codes} (where applicable); \textbf{Competing interests} (mandatory statement). 

% The corresponding author is responsible for submitting a \href{http://www.nature.com/srep/policies/index.html#competing}{competing interests statement} on behalf of all authors of the paper. This statement must be included in the submitted article file. {\color{red}Dr. Wu and Kim, please take care of this.}
\section*{Additional information}

\subsection*{Competing interests}

Dr. Buatti's work has been funded in part by the National Institutes of Health/National Cancer Institute grants P01 CA217797001A1, UL1 TR002537,
U01 CA140206, and 1R21CA209874.
\
Dr. Kim's work has been funded in part by the National Institutes of Health/National Cancer Institute grant 1R21CA209874.
\
Dr. Wu's work has been funded in part by the National Institutes of Health grants 1R21CA209874 and R01EB020665.
\
Drs. Baek, Wu, Kim, Smith, Allen, Buatti and Mr. He are the co-inventors of the provisional patent (U.S. App. No. 62/811,326. Systems And Methods For Image Segmentation And Survival Prediction Using Convolutional Neural Networks) based upon the work of this manuscript.
\
Dr. Plichta, Dr. Seyedin, Ms. Gannon, and Ms. Cabel declare no potential conflict of interest.

\section*{Extended Data}
\captionsetup[table]{name=Extended Data Table}
\captionsetup[figure]{name=Extended Data Figure}
\setcounter{table}{0}
\setcounter{figure}{0}

\begin{table}[ht]
\begin{center}
\caption{Summary statistics of the data set}
\label{tbl:summary_stat}
\small
\begin{tabular}{cccccc}
    \toprule
    \multirow{3}{*}{\makecell{ \\ Age at Diagnosis \\ Age at SBRT}} & {Mean} &{Median} &{Std. Dev.} &{Min}&{Max}\\
    & 70.34 & 71.00 & 10.59 & 47.00 & 90.00 \\
    & 71.90 & 72.36 & 10.58 & 47.13 & 90.17 \\
    \midrule
    \multirow{2}{*}{Group Staging} & {I} &{II} &{III} &{IV}&{Other$^{*}$}\\
    & 37 & 10 & 14 &25 &10
    \\\hline
    \multirow{2}{*}{Smoking Status} & {Non-smoker} &{Quit} &{Light smoker} &{Smoker}\\
    & 10 & 67 & 1 &18
    \\\hline
    \multirow{5}{*}{Survival Status} & & {Type} & {Alive} & {Dead} & {Other$^{**}$}\\
    & \multirow{2}{*}{2yr} & {Overall} & 61 & 35 & 0 \\
    &  & {Disease-specific} & 67 & 25 & 4\\
    & \multirow{2}{*}{5yr} & {Overall} & 26 & 48 & 22 \\
    &  & {Disease-specific} & 58 & 14 & 24\\
    \bottomrule
    \multicolumn{6}{l}{$^{*}$Patients with no staging information.}\\
    \multicolumn{6}{l}{$^{**}$Patients unqualified for the survival category due to the insufficient length of follow-up after initial diagnosis.}
\end{tabular}
\end{center}
\end{table}

\begin{sidewaystable}[ht]
\begin{center}
\caption{Predictive power of linear model on 2-yr./5-yr. survival outcome}
\label{tbl:performance}
\footnotesize
\begin{tabular}{cccccccccccccc}
    \hline
    \multicolumn{2}{c}{
    \multirow{2}{*}{Features}} & \multicolumn{2}{c}{(\textbf{a}) TLG} &\multicolumn{2}{c}{(\textbf{b}) Radiomic\cite{oikonomou2018}} &\multicolumn{2}{c}{(\textbf{c}) Benchmark\cite{hosny2018deep}} &\multicolumn{2}{c}{(\textbf{c}) CT U-Net} &\multicolumn{2}{c}{(\textbf{d}) PET U-Net} &\multicolumn{2}{c}{(\textbf{e}) All U-Net}\\
    & & 2-yr. & 5-yr. &2-yr. & 5-yr. &{2-yr.}  &{5-yr.} &2-yr. & 5-yr. & 2-yr. & 5-yr. & 2-yr. & 5-yr.
    \\\hline
    \multirow{4}{*}{\makecell{Accuracy \\ (95\% CI)}} & \multirow{2}{*}{OS} & 0.53 & 0.55 & 0.62 & 0.65 & {0.69} & {0.75} & {0.84} & {0.86} & {0.83} & {0.87} & {0.86} & {0.89}\\
    && (0.46-0.60) & (0.47-0.63) & (0.54-0.70) & (0.55-0.75) & {(0.65-0.73)} & {(0.70-0.80)} & {(0.81-0.87)} & {(0.80-0.92)} & {(0.79-0.87)} & {(0.80-0.94)} & {(0.79-0.93)} & {(0.85-0.94)}\\
     & \multirow{2}{*}{DS} & 0.57 & 0.57 & 0.64 & 0.80 & {0.73} & {0.76} & {0.86} & {0.89} & {0.87} & {0.88} & {0.88} & {0.88}\\
    && (0.54-0.60) & (0.52-0.62) & (0.59-0.69) & (0.75-0.85) & {(0.67-0.80)} & {(0.69-0.83)} & {(0.80-0.92)} & {(0.84-0.94)} & {(0.81-0.93)} & {(0.83-0.93)} & {(0.81-0.95)} & {(0.82-0.94)}\\
    \hline
    \multirow{4}{*}{\makecell{Sensitivity \\ (95\% CI)}} & \multirow{2}{*}{OS} & {  0.40} & {  0.32} & { 0.44} & { 0.60} & { 0.63} & { 0.69} & { 0.72} & { 0.81} & { 0.75} & { 0.82} & { 0.74} & { 0.82}\\
    && {(0.27-0.53)} & {(0.20-0.44)} & {(0.34-0.54)} & {(0.53-0.67)} & {(0.57-0.70)} & {(0.62-0.76)} & {(0.61-0.83)} & {(0.76-0.88)} & {(0.68-0.82)} & {(0.74-0.90)} & {(0.67-0.81)} & {(0.76-0.88)}\\
     & \multirow{2}{*}{DS} & { 0.17} & { 0.33} & { 0.40} & { 0.91} & { 0.68} & { 0.70} & { 0.77} & { 0.84} & { 0.79} & { 0.85} & { 0.79} & { 0.86}\\
    && {(0.11-0.23)} & {(0.23-0.43)} & {(0.31-0.49)} & {(0.86-0.96)} & {(0.60-0.76)} & {(0.65-0.75)} &
    {(0.67-0.87)} & {(0.78-0.90)} & {(0.70-0.88)} & {(0.78-0.92)} & {(0.73-0.85)} & {(0.80-0.92)}\\
    \hline
    \multirow{4}{*}{\makecell{Specificity \\ (95\% CI)}} & \multirow{2}{*}{OS} & { 0.66} & { 0.78} & { 0.80} & { 0.68} & { 0.77} & { 0.80} & { 0.91} & { 0.90} & { 0.90} & { 0.93} & { 0.91} & { 0.92}\\
    && {(0.55-0.77)} & {(0.71-0.85)} & {(0.68-0.92)} & {(0.60-0.76)} & {(0.70-0.84)} & {(0.75-0.85)} & {(0.78-0.99)} & {(0.83-0.97)} & {(0.80-0.99)} & {(0.84-0.99)} & {(0.84-0.98)} & {(0.84-0.99)}\\
     & \multirow{2}{*}{DS} & { 0.97} & { 0.81} & { 0.88} & { 0.70} & { 0.78} & { 0.81} & { 0.92} & { 0.92} & { 0.90} & { 0.91} & { 0.91} & { 0.91}\\
    && {(0.95-0.99)} & {(0.71-0.91)} & {(0.82-0.94)} & {(0.62-0.78)} & {(0.72-0.84)} & {(0.76-0.86)} & {(0.81-0.99)} & {(0.86-0.98)} & {(0.85-0.95)} & {(0.84-0.98)} & {(0.83-0.99)} & {(0.86-0.96)}\\
    \hline
    \multirow{4}{*}{\makecell{AUC \\ (95\% CI)}} & \multirow{2}{*}{OS} & 0.60 & 0.63 & 0.61 & 0.69 & { 0.70} & { 0.72} & { 0.83} & { 0.85} & { 0.82} & { 0.87} & { 0.88} & { 0.89}\\
    && (0.46-0.74) & (0.56-0.70) & (0.48-0.74) & (0.60-0.78) & {(0.65-0.75)} & {(0.66-0.78)} & {(0.75-0.91)} & {(0.75-0.95)} & {(0.73-0.89)} & {(0.79-0.95)} & {(0.80-0.96)} & {(0.85-0.93)}\\
     & \multirow{2}{*}{DS} & 0.68 & 0.68 & 0.64 & 0.83 & { 0.72} & { 0.73} & { 0.85} & { 0.85} & { 0.86} & { 0.88} & { 0.86} & { 0.88}\\
    && (0.61-0.75) & (0.60-0.76) & (0.57-0.71) & (0.79-0.87) & {(0.67-0.77)} & {(0.68-0.78)} & {(0.75-0.95)} & {(0.77-0.92)} & {(0.79-0.93)} & {(0.80-0.96)} & {(0.81-0.91)} & {(0.81-0.95)}\\
    \hline
\end{tabular}
\end{center}
\end{sidewaystable}

\begin{table}[ht]
\begin{center}
\caption{Predictive power on the extramural data set}
\label{tbl:stanford_stat}
\small
\begin{tabular}{ c c c c c c c c }
    \hline
    \multicolumn{2}{c}{
    \multirow{2}{*}{Features}} &\multicolumn{2}{c}{(\textbf{a}) Radiomics$^*$} &\multicolumn{2}{c}{(\textbf{b}) Benchmark\cite{hosny2018deep}} &\multicolumn{2}{c}{(\textbf{c}) CT U-Net} \\
    & &{2-yr.}  &{5-yr.} &{2-yr.}  &{5-yr.} &{2-yr.}  &{5-yr.} 
    \\\hline
    \multirow{2}{*}{\makecell{Accuracy \\ (95\% CI)}} & \multirow{2}{*}{OS} & { 0.46} & {0.71} & {0.60} & {0.65} & {0.82} & {0.87}\\
    && {(0.34-0.58)} & {(0.62-0.80)} & {(0.50-0.70)} & {(0.53-0.77)} & {(0.77-0.87)} & {(0.80-0.94)}\\
    \hline
    \multirow{2}{*}{\makecell{Sensitivity \\ (95\% CI)}} & \multirow{2}{*}{OS} & {0.38} & {0.63} & {0.40} & {0.45} & {0.75} & {0.90} \\
    && {(0.30-0.46)} & {(0.52-0.74)} & {(0.33-0.47)} & {(0.40-0.50)} & {(0.65-0.85)} & {(0.80-0.99)}\\
    \hline
    \multirow{2}{*}{\makecell{Specificity \\ (95\% CI)}} & \multirow{2}{*}{OS} & {0.59} & {0.74} & {0.83} & {0.88} & {0.91} & {0.85}\\
    && {(0.51-0.67)} & {(0.67-0.81)} & {(0.72-0.94)} & {(0.80-0.96)} & {(0.87-0.95)} & {(0.81-0.89)}\\
    \hline
    \multirow{2}{*}{\makecell{AUC \\ (95\% CI)}} & \multirow{2}{*}{OS} & {0.49} & {0.75} & {0.65} & {0.71} & {0.87} & {0.90}\\
    && {(0.40-0.58)} & {(0.67-0.83)} & {(0.57-0.73)} & {(0.60-0.82)} & {(0.80-0.94)} & {(0.82-0.98)}\\
    \hline
    \multicolumn{8}{l}{$^{*}$The extramural data set does not have PET images, hence, only 6 features out of the original 17 features}\\
    \multicolumn{8}{l}{\, in Oikonomou \etal\cite{oikonomou2018} were included.}\\
\end{tabular}
\end{center}
\end{table}

\begin{figure}[!]
\centering
\includegraphics[width=0.9\linewidth]{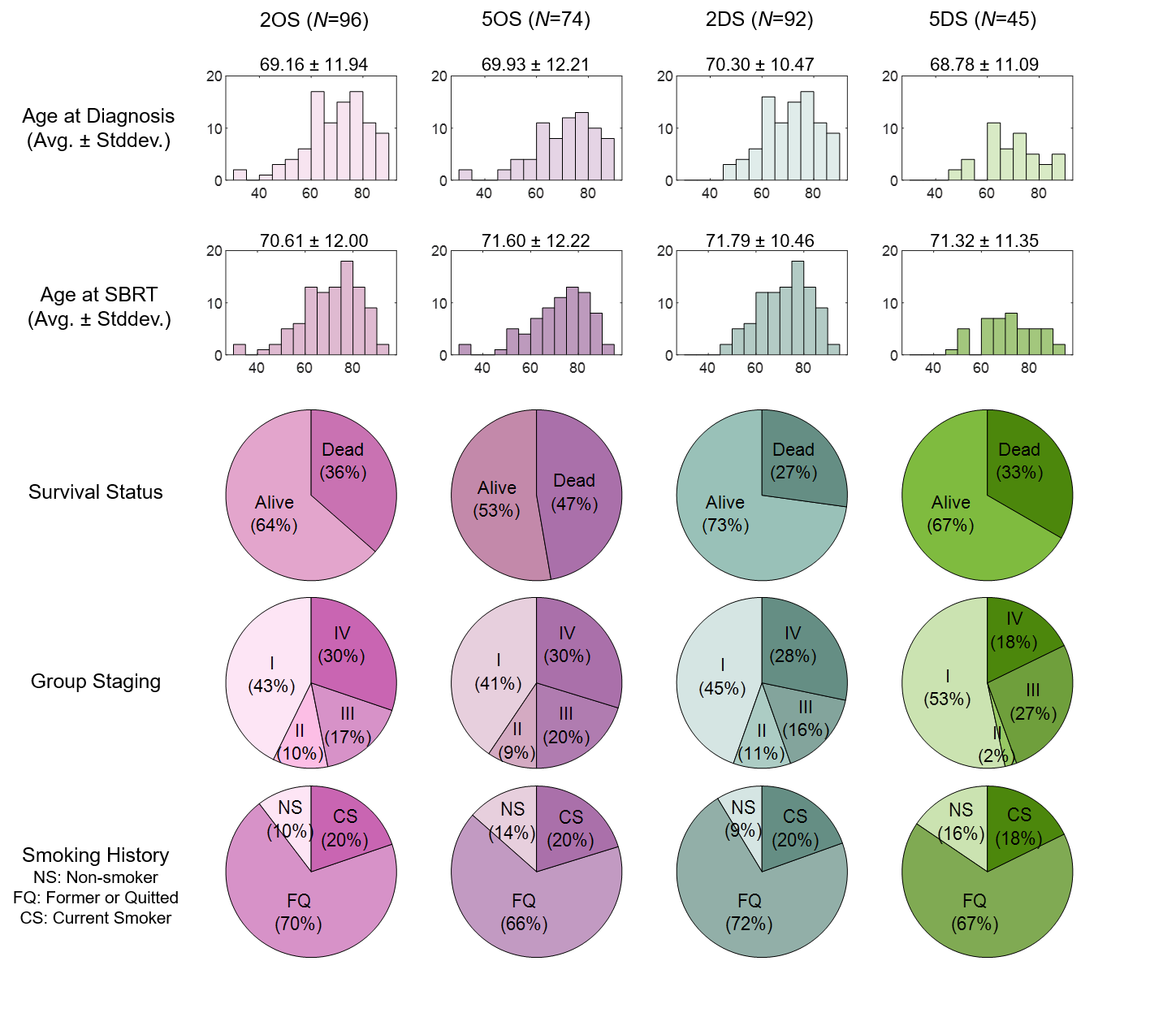}
\caption{Summary statistics of the patients utilized in this study}
\label{fig:Summary_stat}
\end{figure}

\begin{figure}[!]
\centering
\includegraphics[width=\linewidth]{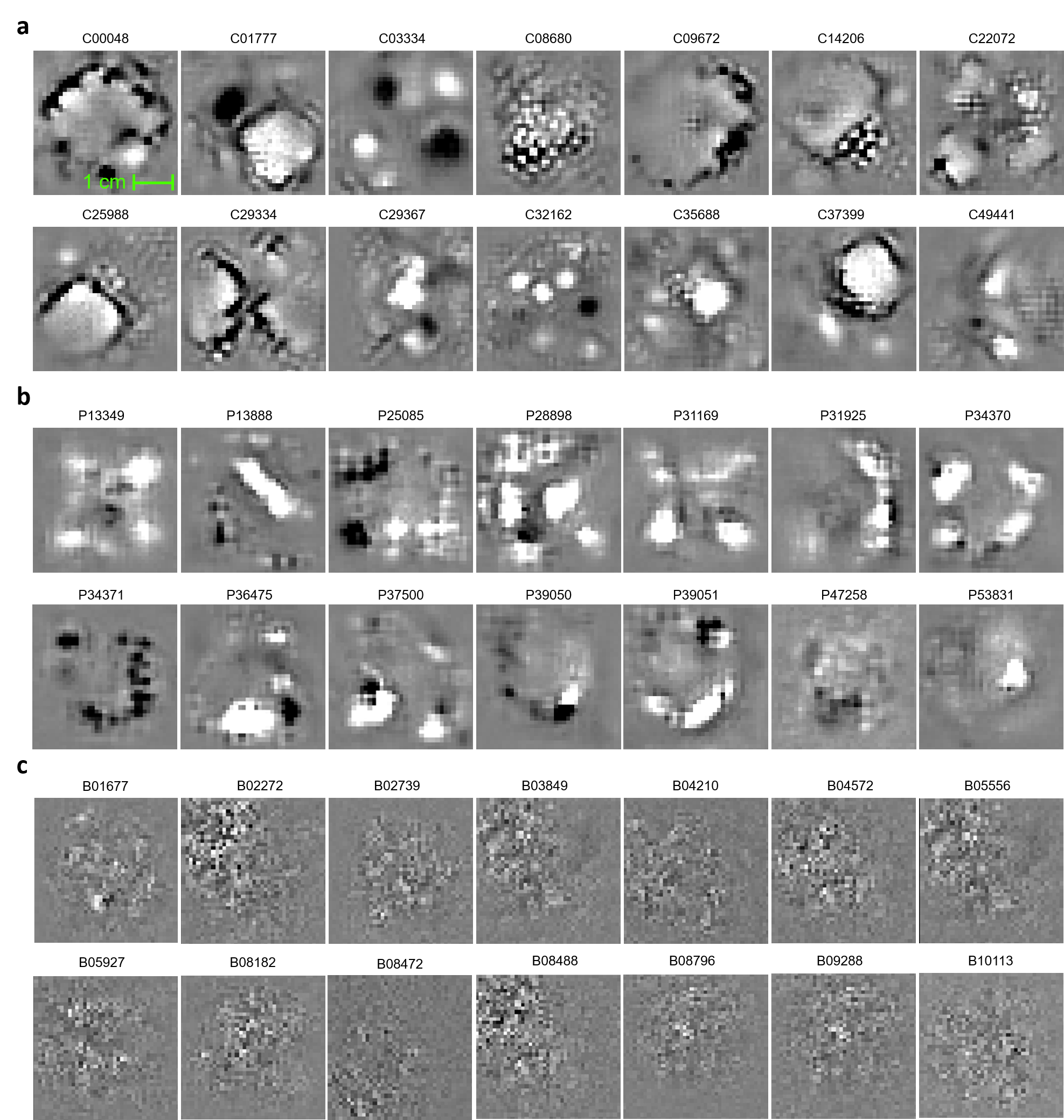}
\caption{Visualization of survival-related features captured by CNNs}
\label{fig:activation_maximization_full}
\end{figure}

\begin{figure}[!]
\centering
\includegraphics[width=0.65\linewidth]{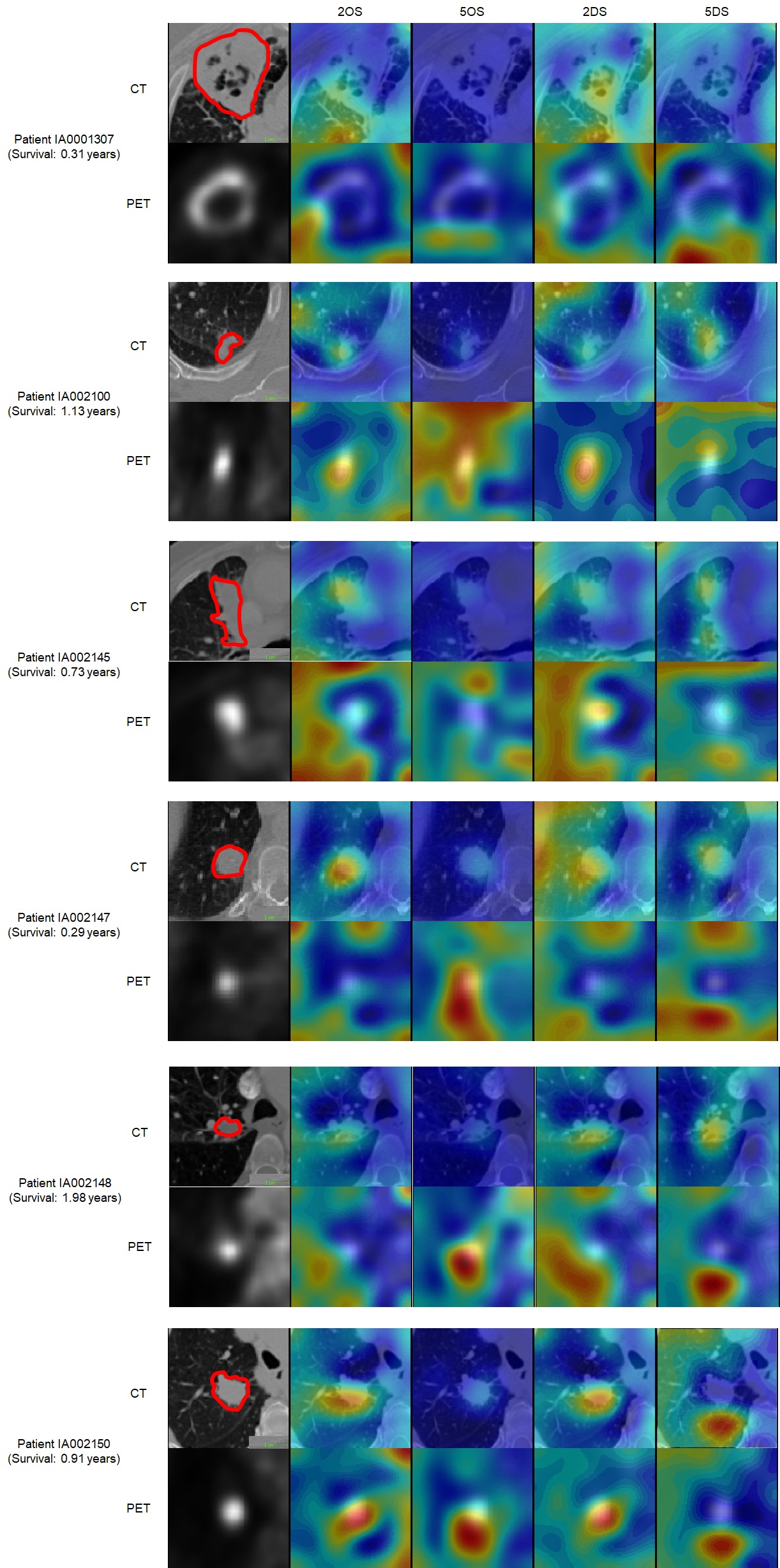}
\caption{Risk map visualization of the patients under 2 years of survival.}
\label{fig:riskmap_under2}
\end{figure}

\begin{figure}[!]
\centering
\includegraphics[width=0.65\linewidth]{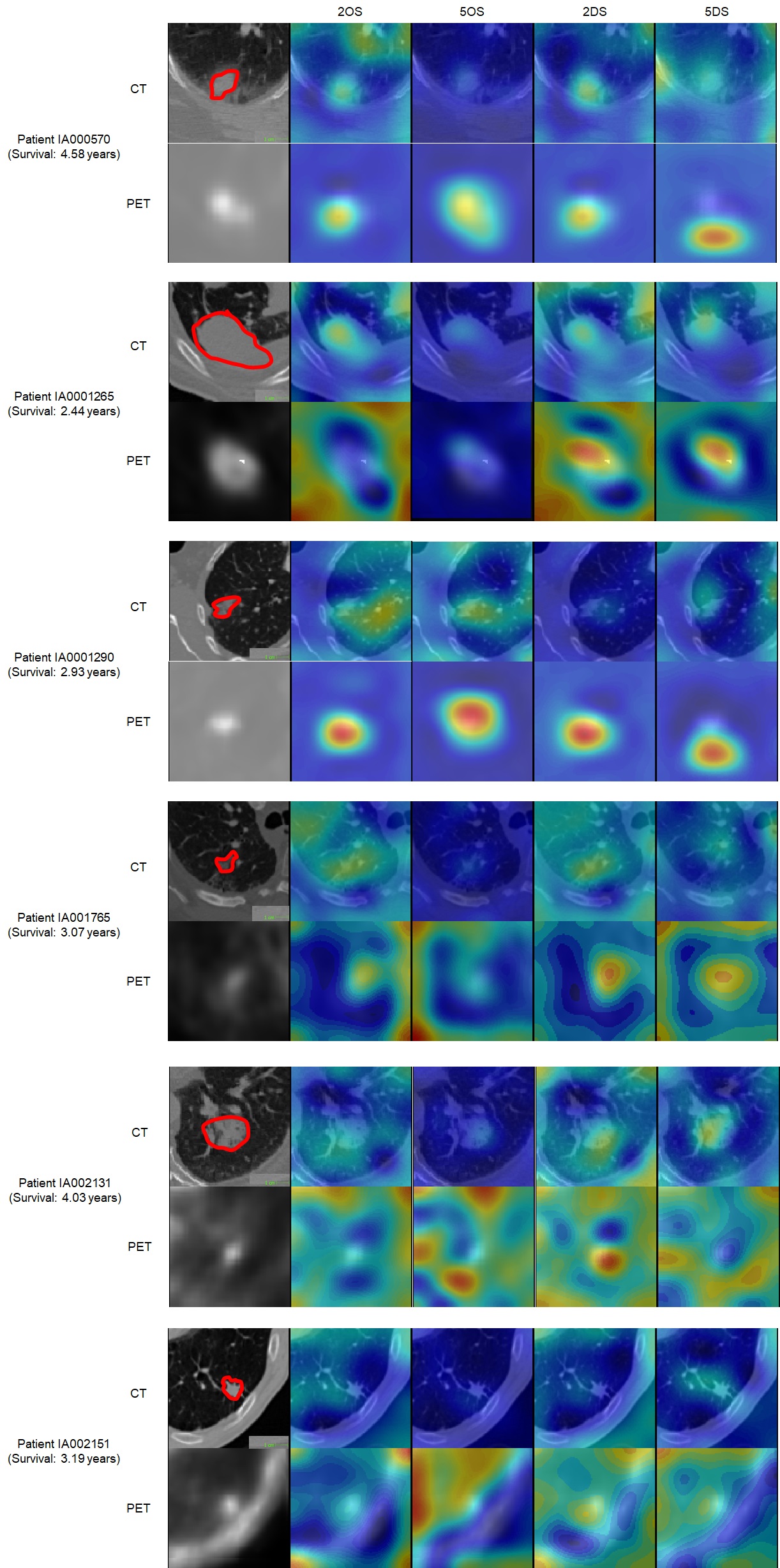}
\caption{Risk map visualization of the patients between 2 and 5 years of survival.}
\label{fig:riskmap_2to5}
\end{figure}

\begin{figure}[!]
\centering
\includegraphics[width=0.65\linewidth]{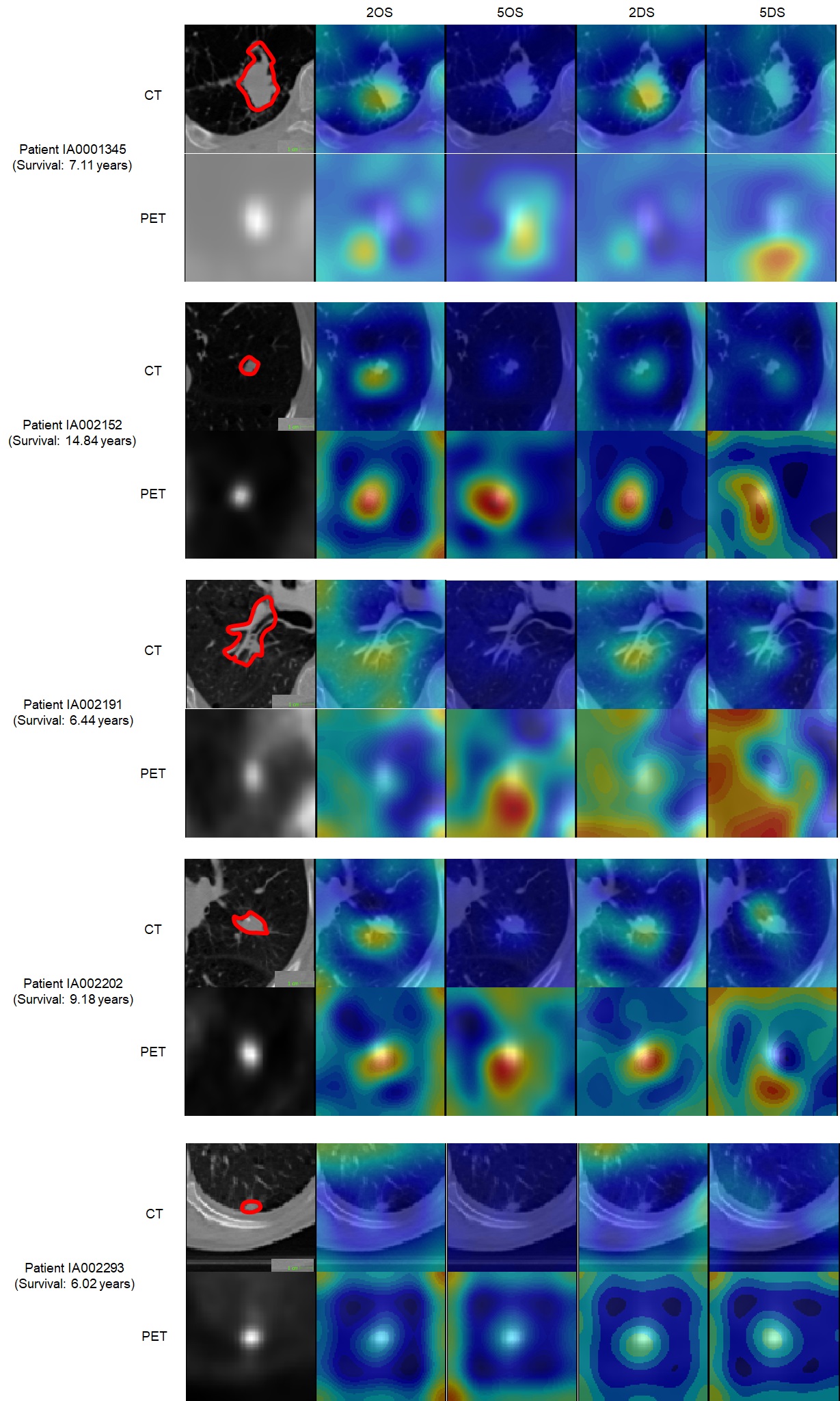}
\caption{Risk map visualization of the patients over 5 years of survival.}
\label{fig:riskmap_over5}
\end{figure}

\end{document}